\documentclass[sigconf]{acmart}

\acmConference[ASE '26]{39th IEEE/ACM International Conference on Automated Software Engineering}{October 12--16, 2026}{Munich, Germany}
\acmYear{2026}
\copyrightyear{2026}
\setcopyright{none}
\acmDOI{}
\acmISBN{}

\newcommand{\vInv}{\mathit{vInv}}
\newcommand{\sInv}{\mathit{sInv}}

\definecolor{findingbg}{RGB}{235,235,235}
\newcommand{\finding}[2]{%
  \smallskip\noindent
  \colorbox{findingbg}{\parbox{\dimexpr\columnwidth-2\fboxsep}{%
    \smallskip\textbf{#1}\enspace #2\smallskip}}%
  \smallskip}

\usepackage{listings}
\usepackage{stmaryrd}

\definecolor{rustkw}{RGB}{160,32,240}
\definecolor{rustcmt}{RGB}{0,128,0}
\definecolor{ruststr}{RGB}{163,21,21}
\definecolor{kanikw}{RGB}{0,0,200}

\lstdefinelanguage{Rust}{
  keywords={fn, pub, impl, unsafe, let, mut, as, self, use, mod, struct,
    enum, trait, where, for, in, match, if, else, loop, while, break,
    continue, return, crate, super, true, false, static},
  keywordstyle=\color{rustkw}\bfseries,
  ndkeywords={u8, u32, u64, usize, bool, Result, Option, Vec, Some, None},
  ndkeywordstyle=\color{rustkw},
  comment=[l]{//},
  commentstyle=\color{rustcmt}\ttfamily,
  stringstyle=\color{ruststr}\ttfamily,
  morestring=[b]",
  sensitive=true
}

\title{Kani: A Model Checker for Rust}

\author{R\'emi Delmas, Zyad Hassan, Qinheping Hu, Rahul Kumar, Felipe R. Monteiro, Thanh Nguyen, Adri\'an Palacios}
\affiliation{\institution{Amazon Web Services}\city{Seattle}\state{WA}\country{USA}}

\author{Celina Val}
\affiliation{\institution{Amazon Web Services}\city{Vancouver}\state{BC}\country{Canada}}

\author{Michael Tautschnig}
\affiliation{\institution{Amazon Web Services}\city{Seattle}\state{WA}\country{USA}}
\affiliation{\institution{Queen Mary University}\city{London}\country{United Kingdom}}

\author{Justus Adam}
\authornote{All work done while at Amazon Web Services.}
\affiliation{\institution{Brown University}\city{Providence}\state{RI}\country{USA}}

\author{Daniel Schwartz-Narbonne}
\authornotemark[1]
\affiliation{\institution{Datadog}\city{New York}\state{NY}\country{USA}}

\author{Carolyn Zech}
\authornotemark[1]
\affiliation{\institution{Massachusetts Institute of Technology}\city{Cambridge}\state{MA}\country{USA}}

\renewcommand{\shortauthors}{Delmas et al.}

\begin{document}




\begin{abstract}
Rust's ownership type system prevents memory errors in safe code,
but certain desirable properties remain orthogonal to compilation:
the soundness of \texttt{unsafe} operations (e.g., raw pointer
dereferences),
functional correctness, and absence of runtime panics.  We present Kani, an
open-source model checker for Rust that pushes bounded model
checking beyond bug-finding to provide correctness guarantees for
these properties.  Kani compiles proof harnesses from Rust's Mid-level
Intermediate Representation (MIR) into CBMC's bit-precise verification
engine, automatically checking a comprehensive set of safety properties with
no user annotation.  To extend verification from bounded to
unbounded, Kani provides a specification language comprising
function contracts, loop contracts, quantifiers, and function
stubbing.  We demonstrate feasibility through case studies on
industrial Rust projects, where contracts upgraded verification from
panic-freedom to functional correctness, uncovering six previously
unknown bugs.  Kani operates at scale in production CI, with over
16,000 harnesses verified per code change in the Rust standard
library verification campaign.
\end{abstract}

\maketitle

\section{Introduction}
\label{sec:introduction}

Rust has become the language of choice for safety-critical systems
software, from embedded operating
systems~\cite{Levy17,Boos20} and cryptographic
protocols~\cite{Bhargavan25} to cloud
infrastructure~\cite{Agache20}.  Its ownership type system
statically prevents data races and many classes of memory
errors~\cite{Jung18}, but there are three classes of correctness
properties that the compiler cannot prove:
(1)~the correctness of \texttt{unsafe} operations (e.g., raw pointer
dereferences, calls to \texttt{unsafe} functions, access to mutable
statics, \texttt{unsafe} trait implementations, and \texttt{union}
field access) where the compiler still enforces borrow checking and
type safety but the developer assumes responsibility for the
additional safety invariants these operations
require~\cite{Astrauskas20unsafe,Cui24achilles};
(2)~functional correctness properties such as algorithmic
correctness and protocol conformance; and
(3)~absence of runtime panics from operations like
\texttt{unwrap()}, integer overflow, and out-of-bounds access, in
contexts where panics are undesirable.  Testing and fuzzing can expose
some of these failures, but they explore only a finite sample of the
input space and provide no completeness guarantee.

Deductive verification tools for Rust, including
Prusti~\cite{Astrauskas22}, Creusot~\cite{Denis22}, and
Verus~\cite{Lattuada23verus}, can prove rich functional properties.
However, they require substantial proof
engineering (e.g., separation logic or ghost state)
that limits adoption outside specialist
teams~\cite{Huang26}.  At the other end of the spectrum, dynamic
tools like Miri~\cite{Jung26miri} detect undefined behavior at
runtime but cannot prove its absence.  There is a gap between these
extremes: developers need a verification approach that can start
with high automation and low annotation cost, then scale
incrementally toward stronger correctness guarantees as the
verification need grows.

Bounded model checking fills the automated end of this gap.  A
bounded model checker encodes default safety properties (e.g., arithmetic
overflow, division by zero, null dereferences, assertion violations)
and checks them exhaustively up to a bound, requiring minimal manual
proof construction.  This makes it an effective entry point for
verification: developers write proof harnesses that resemble unit
tests, and the tool proves properties over all inputs within the
bound.  The limitation is that bounded analysis alone cannot
guarantee correctness beyond the unwinding depth.  To push model
checking from bug-finding toward correctness proofs, the technique
must be extended with specification constructs that enable unbounded
reasoning while preserving the low annotation overhead that makes
bounded checking accessible.

This paper presents Kani, an open-source model checker for Rust that
realizes this progression from bounded analysis to unbounded
correctness guarantees.  Kani builds on the code-level model
checking methodology~\cite{Chong20,Chong21}, which demonstrated
that bounded model checking with CBMC~\cite{cbmc14} can be
integrated into continuous development workflows for industrial C
codebases.  Kani extends this methodology to Rust, operating on the
Mid-level Intermediate Representation (MIR) to preserve
Rust-specific type invariants, and adds a specification language
comprising function contracts, loop contracts, quantifiers, and
function stubbing.  These constructs allow developers to
incrementally annotate their code: Kani first proves default safety
properties with no annotation, then, as contracts are added, extends
the same verification engine to unbounded functional correctness
proofs.  We demonstrate the feasibility of this approach through a
case study on the Hifitime time-management library, where adding
contracts upgraded guarantees from panic-freedom to functional
correctness at low specification overhead, with AI-assisted
specification drafting via an AI coding assistant.  Kani is
deployed in production CI at scale: proof harnesses run on every
code change in Firecracker~\cite{Agache20}, the virtual machine
monitor behind AWS Lambda and AWS Fargate; in s2n-quic, Amazon's
Rust implementation of the IETF QUIC transport protocol; in
Hifitime~\cite{Hifitime}, a time-management library used in
aerospace applications; and in the
Rust standard library verification
campaign~\cite{Cook25verifyrust}, where over 16,000 harnesses are
verified per code change.

We make the following contributions:
\begin{enumerate}
\item We evaluate Kani on multiple industrial Rust projects, where it
  uncovered eleven bugs missed by testing and fuzzing.  Through a
  detailed case study on the Hifitime library, we show that
  contract-based verification extends Kani's default safety checks
  to unbounded functional correctness proofs at low
  specification overhead (\S\ref{sec:eval}).
\item We formalize Kani's specification language for function
  contracts, loop contracts, quantifiers, and function stubbing,
  grounding the semantics in Floyd-Hoare logic and connecting them
  to the underlying bounded model checking engine
  (\S\ref{sec:methodology}).
\item We present Kani's architecture, its MIR-level design, and its
  integration with the Rust toolchain
  (\S\ref{sec:architecture}).
\end{enumerate}

\section{Kani by Example}
\label{sec:example}

We illustrate Kani's verification workflow using Euclid's greatest
common divisor (GCD) algorithm and its use in
Firecracker~\cite{Agache20},%\footnote{\url{https://github.com/firecracker-microvm/firecracker}}
an open-source virtual machine monitor that runs workloads in
lightweight microVMs.  Firecracker powers AWS Lambda and AWS
Fargate, making the correctness of its Rust codebase a
security-critical concern.  The GCD function is used in
Firecracker's rate limiter to simplify token-bucket refill ratios,
and its iterative loop makes it a natural target for both bounded
and unbounded verification.

\subsection{Bounded Verification}
\label{sec:bounded-example}

Consider the iterative GCD implementation from Firecracker's
\texttt{rate\_limiter} module:
\begin{lstlisting}
fn gcd(x: u64, y: u64) -> u64 {
    let mut a = x;
    let mut b = y;
    while b != 0 {
        let t = b;
        b = a % b;
        a = t;
    }
    a
}
\end{lstlisting}
A developer can write a Kani \emph{proof harness}, analogous to a
unit test but over all possible inputs, to verify that \texttt{gcd}
returns a common divisor:
\begin{lstlisting}
#[kani::proof]
#[kani::unwind(94)]
fn check_gcd() {
    let x: u64 = kani::any();
    let y: u64 = kani::any();
    kani::assume(x > 0 && y > 0);
    let d = gcd(x, y);
    assert!(d != 0 && x % d == 0 && y % d == 0);
}
\end{lstlisting}
The call \texttt{kani::any()} produces a nondeterministic value of
the given type, and \texttt{kani::assume} constrains the input
space.  A note of caution: an incorrect assumption (e.g.,
\texttt{assume(false)}) makes the proof vacuously true, so
assumptions must be reviewed as carefully as the code itself.  Kani unwinds the loop, converts the result to SSA form, and
encodes all assertions as a propositional formula
(\S\ref{sec:bmc}) that is checked using a
SAT solver.

The \texttt{\#[kani::unwind(94)]} annotation sets the loop
unwinding bound.  For 64-bit inputs, the worst-case number of
iterations of Euclid's algorithm is 93 (the largest Fibonacci
number below $2^{64}$ is $F_{93}$); the bound is set conservatively
to~94.  If the bound is
insufficient, Kani reports an \emph{unwinding assertion failure},
alerting the developer that the verification result is inconclusive
beyond the given depth.  Even with a sufficient bound, the resulting
formula is very large: in our experiments, bounded verification of
\texttt{gcd} over the full \texttt{u64} range did not terminate
within a one-hour timeout.  Moreover, the bound is specific to
64-bit integers; changing the input type requires recomputing it.

\subsection{Unbounded Verification with Contracts}
\label{sec:unbounded-example}

We annotate \texttt{gcd} with a function contract (precondition and
postcondition) and a loop contract (invariant and decreases clause).
The loop contract eliminates the dependence on the unwinding bound
by abstracting the loop via an inductive invariant:
\begin{lstlisting}
#[kani::requires(x > 0 && y > 0)]
#[kani::ensures(|&result| result > 0)]
fn gcd(x: u64, y: u64) -> u64 {
    let mut a = x;
    let mut b = y;
    #[kani::loop_invariant(
        a > 0
        && kani::forall!(|d: u64
            in (1, a.saturating_add(1))|
            d == 0
            || ((x % d == 0 && y % d == 0)
            == (a % d == 0 && b % d == 0)))
    )]
    #[kani::loop_decreases(b)]
    while b != 0 {
        let t = b;
        b = a % b;
        a = t;
    }
    a
}
\end{lstlisting}
The \texttt{requires} clause states the precondition: both inputs
must be positive.  The \texttt{ensures} clause states the
postcondition: the result is positive.  Inside the loop,
\texttt{loop\_invariant} asserts two properties across all
iterations: (1)~\texttt{a} remains positive, and (2)~for every
candidate divisor~\texttt{d}, \texttt{d}~divides the original
inputs \texttt{(x,\,y)} iff it divides the current values
\texttt{(a,\,b)}, i.e., the set of common divisors is preserved.
The \texttt{saturating\_add} avoids overflow when computing the
upper bound of the quantifier range.  The \texttt{loop\_decreases}
clause specifies that \texttt{b} is a well-founded decreasing
measure (since $a \bmod b < b$ on each iteration), proving
termination (\S\ref{sec:loop-contracts}).

\noindent
The natural postcondition is the full divisibility property
(\texttt{x \% result == 0 \&\& y \% result == 0}), which the
invariant implies at loop exit; however, Z3 cannot automatically
instantiate the quantifier at $d = \mathit{result}$ over nonlinear
64-bit bitvector remainder (\S\ref{sec:quantifiers}), so we use the
weaker \texttt{result > 0}.  Verification passes all 202 checks in
0.54\,s with Z3, exercising the full pipeline: quantified loop
invariant, termination proof, and all automatically generated safety
checks.

This invariant is nontrivial: it requires a universally quantified
statement over nonlinear arithmetic.  In practice, many verification
tasks require only simple contracts, which Kani checks automatically
with no annotations.  The GCD example
demonstrates the expressiveness of Kani's specification language
when functional correctness is desired.

\paragraph{Verifying the contract.}
A dedicated \texttt{proof\_for\_contract} harness sets nondeterministic
inputs via \texttt{kani::any()} and calls \texttt{gcd}; Kani assumes
the precondition, executes the function, and asserts the postcondition.
The \texttt{\#[kani::solver(z3)]} attribute selects the Z3 SMT solver,
required for the quantified invariant.
The loop invariant is verified inductively
(\S\ref{sec:loop-contracts}): the loop body executes exactly twice
in a single BMC query, regardless of input size.

\paragraph{What failure looks like.}
When a postcondition is violated, Kani reports the contract
expression and source location.  For example, a sign error in
Hifitime's \texttt{total\_nanoseconds()} (\S\ref{sec:hifitime})
produced:
\begin{lstlisting}[language={},basicstyle=\ttfamily\scriptsize]
Check 1: ...total_nanoseconds::{closure#2}
 - Status: FAILURE
 - Description: "|result| { *result ==
     i128::from(self.centuries) *
     i128::from(NPC) +
     i128::from(self.nanoseconds) }"
VERIFICATION:- FAILED (0.26s)
\end{lstlisting}

\paragraph{Using the verified contract.}
Once verified, the contract serves as a sound abstraction for
compositional reasoning.  In Firecracker,
\texttt{TokenBucket::new} calls \texttt{gcd} to simplify
token-bucket refill ratios.  Using
\texttt{\#[kani::stub\_verified(gcd)]}, the harness replaces
\texttt{gcd} with its contract: the precondition is asserted at the
call site, and a nondeterministic return value satisfying the
postcondition (\texttt{result > 0}) is assumed.  No loop is
unrolled, and verification completes in seconds over all 64-bit
inputs using a sound abstraction.

\section{Architecture}
\label{sec:architecture}

\begin{figure*}[t]
  \centering
  \includegraphics[width=\textwidth]{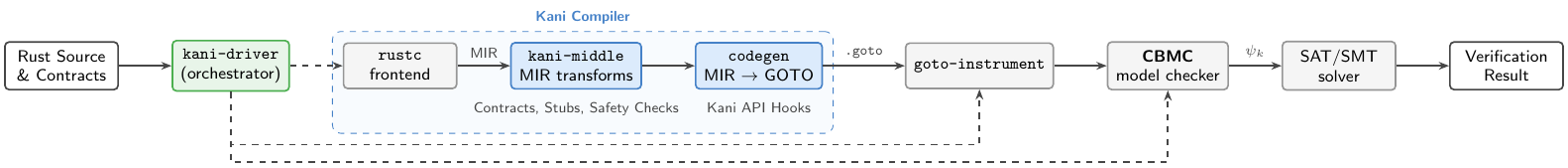}
  \caption{Kani verification pipeline.  Blue boxes are Kani
    components; gray boxes are external tools; green is the
    orchestrator.  White boxes denote input/output.
    Dashed arrows indicate orchestration control flow.}
  \label{fig:arch}
\end{figure*}

Kani is an open-source verification tool that integrates with the
standard Rust toolchain.  It can be invoked via \texttt{cargo~kani}
on Cargo\footnote{\url{https://doc.rust-lang.org/cargo/}} packages
(analogous to \texttt{cargo~test}) or via \texttt{kani} on
individual crates.  Internally, Kani reuses the entire
\texttt{rustc} frontend and replaces the code-generation backend:
instead of emitting machine code via LLVM, it translates Rust
programs into GOTO programs~\cite{Kroening23} that CBMC can model-check.  A key design
decision is to operate on Rust's Mid-level Intermediate
Representation (MIR) rather than LLVM IR.  MIR preserves
monomorphized type information, enum discriminant layouts, and
Rust's validity invariants (e.g., that a \texttt{bool} is 0 or 1,
that a \texttt{char} is a valid Unicode scalar) in a form that LLVM
IR discards.  This enables Kani to automatically check
Rust-specific safety properties and to reason about dynamic trait
objects and closures, which are erased at the LLVM
level~\cite{VanHattum22}.  As shown in Figure~\ref{fig:arch}, the
pipeline flows from Rust source through the Kani compiler to CBMC
and a SAT/SMT solver, with the \textit{kani-driver} orchestrating each
stage.

\paragraph{Compilation.}
A verification session starts with \texttt{cargo kani}, which
invokes \emph{kani-driver}, the orchestrator (green box in
Figure~\ref{fig:arch}).  The driver calls \emph{kani-compiler}, a
\texttt{rustc} plugin that hooks into the compiler via the
\texttt{rustc\_private} interface after MIR
generation.  The standard \texttt{rustc} frontend performs parsing,
name resolution, type checking, trait resolution, and
monomorphization, producing typed MIR.  Kani then applies
MIR-to-MIR transformations: reachability analysis, contract and
loop-contract instrumentation (\S\ref{sec:methodology}), stubbing,
and safety-check insertion.

\paragraph{Code generation.}
The transformed MIR is lowered to a GOTO program by the
\emph{codegen} backend, which translates MIR statements, rvalues,
places, and types into GOTO expressions using the
\emph{cprover-bindings} library available in CBMC.  Calls to the
Kani API (e.g., \texttt{kani::any()}, \texttt{kani::assume()},
\texttt{kani::assert()}, \texttt{forall!}, or \texttt{exists!}) are
intercepted by codegen hooks and translated into their CBMC
counterparts: nondeterministic values, assumptions, assertions, and
quantified expressions.  Quantifier bodies are inlined to produce
single CBMC-compliant expressions (\S\ref{sec:quantifiers}).  The
resulting GOTO program is serialized to CBMC's binary IRep format.

\paragraph{Verification.}
The driver optionally invokes \emph{goto-instrument} to apply
Dynamic Frame Condition
Checking\footnote{\url{https://diffblue.github.io/cbmc/contracts-dev-spec-dfcc.html}}
(DFCC) for function and loop contracts, instrumenting write-set
enforcement.  It then invokes CBMC, which performs symbolic
execution of the GOTO program, unwinds loops, converts the result
to SSA form, and encodes all assertions as a propositional formula
(\S\ref{sec:bmc}), which it dispatches to a solver.  Kani supports
multiple solver backends:
MiniSat~\cite{Een03} (default), Kissat and
CaDiCaL~\cite{Biere20} for SAT, and Z3~\cite{Moura08},
cvc5~\cite{Barbosa22}, and Bitwuzla~\cite{Niemetz23} for SMT.
CBMC returns a verdict per property (e.g., assertion, overflow
check, memory safety check), which the driver maps back to Rust
source locations and presents to the user.

%\paragraph{CI integration.}
%Kani is designed to fit into continuous integration workflows.
%Since \texttt{cargo kani} returns a nonzero exit code on any
%verification failure, it can be added as a CI step alongside
%\texttt{cargo test}.  Proof harnesses live in the same source tree
%as the code under verification (typically gated behind
%\texttt{\#[cfg(kani)]}), so they are version-controlled and
%reviewed together with the code they verify.  This follows the
%code-level model checking methodology~\cite{Chong20,Chong21}, where
%proofs are maintained as part of the development workflow rather
%than as a separate verification effort.

%% Commented out for space — full property table
%\begin{table}[htbp]
%\centering
%\caption{Automatically checked properties.  Source: Kani
%  codegen~(K) or CBMC symbolic execution~(C).}
%\label{tab:checks}
%\footnotesize
%\begin{tabular}{@{}lcl@{}}
%\hline
%\textbf{Property} & \textbf{Src} & \textbf{Note} \\
%\hline
%Signed overflow        & K & Replaces CBMC's C check \\
%Unsigned overflow      & C & \\
%Division by zero       & K & \\
%Undefined shift        & K & Replaces CBMC's C check \\
%Null pointer deref     & C & \\
%Dangling pointer deref & C & \\
%Misaligned ptr-to-ref  & K & \\
%Array out-of-bounds    & K+C & Both active; Rust fires first \\
%Panics / \texttt{unwrap} & K & \\
%Unreachable code       & K & \\
%Float-to-int finiteness & K & \\
%Exact division remainder & K & \\
%\hline
%Valid value invariants & K & Opt-in \\
%Uninit.\ memory access & K & Opt-in \\
%\hline
%\end{tabular}
%\end{table}

\paragraph{Default properties.}
Beyond user-specified assertions, Kani automatically verifies a
comprehensive set of
properties\footnote{Full list at \url{https://github.com/model-checking/kani}.}
without any annotation from the developer.  These fall into three
categories: \emph{absence of undefined behavior} (i.e., invalid or dangling pointer
dereferences, misaligned casts, invalid enum discriminants),
\emph{absence of runtime panics} (e.g., arithmetic overflow, division by zero,
undefined shift, out-of-bounds indexing, \texttt{unwrap()} on
\texttt{None}), and \emph{resource properties} (e.g., float-to-integer
cast finiteness).  The Kani compiler instruments the code during
MIR transformations and code generation with Rust-semantic checks;
CBMC contributes additional C-level memory safety checks.
Two optional MIR passes provide deeper checks: the
\emph{valid-value pass} asserts that unsafe type coercions produce
valid inhabitants, and the \emph{uninit-memory pass} uses shadow
memory to verify that every dereference accesses initialized memory.
Both are opt-in due to solver overhead.
%% Commented out for space — detail already covered by Default properties paragraph
%Kani-inserted checks are
%generated during code generation: the backend intercepts MIR
%operations and emits assertions for overflow, division by zero,
%undefined shifts, pointer alignment, and every Rust
%\texttt{panic!}/\texttt{assert!}/\texttt{unwrap()}.
%CBMC contributes additional checks during
%symbolic execution, including null and dangling pointer
%dereferences and array bounds violations.  Kani's driver filters
%redundant checks and maps all verdicts back to Rust source
%locations.

\section{Specification Language and Verification Methodology}
\label{sec:methodology}

Kani's verification model follows a deliberate progression: bounded model
checking provides push-button safety analysis for any Rust program; loop and
function contracts extend this to \emph{unbounded} verification when loops or
recursion make a finite bound insufficient.  This section formalizes that
progression, beginning with the bounded foundation, then showing how each
specification construct extends it.  Rust's ownership model introduces
subtleties, particularly around type invariants and memory
havocking, that do not arise in C-level model checking; we identify these
precisely and explain how Kani addresses them.  Throughout, we ground the
formalism in the GCD--\texttt{TokenBucket} example of Section~\ref{sec:example}.

% ---------------------------------------------------------------------------
\subsection{Bounded Model Checking Foundations}
\label{sec:bmc}
% ---------------------------------------------------------------------------

Kani's verification engine is CBMC~\cite{cbmc14,Kroening23}, a bit-precise
bounded model checker (BMC).  Unlike classical BMC formulations defined over
explicit transition systems~\cite{Biere99bmc}, CBMC operates directly on
programs: given a loop \texttt{while}~$G$~$\{B\}$ with guard~$G$ and
body~$B$, CBMC syntactically unwinds it up to a bound~$k$, inserting an
\emph{unwinding assertion} $\neg G$ after the last copy.  The result is a
loop-free program~$P_k$.  CBMC then converts~$P_k$ to Static Single
Assignment (SSA) form and encodes all assertions as a propositional
formula~\cite{Chong21,Kroening03}.  If the formula is satisfiable, the
satisfying assignment yields a concrete counterexample; if unsatisfiable,
all assertions hold for every execution of~$P_k$.

\paragraph{SSA encoding.}
CBMC converts $P_k$ to SSA form, introducing a fresh variable
version~$x_{(i)}$ at each assignment point~$i$ and encoding
branching with path-condition guards~$g_j$.  The result is a
conjunction of guarded equations where $e_j$ is the right-hand-side
expression at assignment~$j$:
\[
  \mathit{SSA}(P_k) \;=\; \bigwedge_{j}\;
  \bigl(g_j \;\rightarrow\; x_{n_j} = e_j\bigr).
\]
The propositional formula CBMC constructs is:
\begin{equation}\label{eq:bmc}
  \psi_k \;=\; \mathit{SSA}(P_k) \;\wedge\;
  \bigvee_{p \,\in\, \mathit{Asserts}(P_k)} \neg\,\varphi_p
\end{equation}
where $\mathit{Asserts}(P_k)$ includes user assertions, automatically
inserted safety checks, and unwinding assertions.

\paragraph{Completeness.}
The bounded guarantee becomes practically complete when all unwinding
assertions pass, meaning $P_k$ covers all reachable paths.  For
programs with input-independent loop bounds this is tractable: in our
running example, Euclid's algorithm iterates at most 93 times, so
\texttt{\#[kani::unwind(94)]} is a sound, complete bound, but the
unwound formula ($O(k \cdot |\mathit{body}|)$ SSA equations over
64-bit bitvectors) exceeds practical solver limits.  For programs
with input-dependent bounds, no tractable~$k$ may suffice.  Function
and loop contracts (\S\ref{sec:loop-contracts}) address this gap by
replacing exhaustive unrolling with inductive reasoning, achieving
unbounded guarantees with a single BMC query.

Kani compiles Rust's MIR into CBMC's GOTO program representation, inheriting
bit-precise arithmetic and memory reasoning.  In the remainder of this
section, we use $\sigma$ for program states and standard Hoare-logic
notation.  The specification constructs below extend this foundation.

% ---------------------------------------------------------------------------
\subsection{Specification Language}
\label{sec:spec-lang}
% ---------------------------------------------------------------------------

Kani's specification language extends Rust boolean expressions with
contract annotations.  All specification expressions must be deterministic
and side-effect free (i.e., no I/O, mutation, or heap allocation), a restriction
enforced by a MIR traversal at compile time.

\paragraph{Syntax.}
The grammar of contract annotations is:
\begin{center}\footnotesize
\begin{tabular}{rcl}
\textit{func-spec} &$::=$&
  $\texttt{requires}(P) \mid \texttt{ensures}(|r|\;Q)$ \\
  && $\mid\; \texttt{modifies}(e)$ \\[3pt]
\textit{loop-spec} &$::=$&
  $\texttt{loop\_invariant}(I)$ \\
  && $\mid\; \texttt{loop\_modifies}(e)
   \mid \texttt{loop\_decreases}(d)$ \\[3pt]
$P, Q, I$ &$::=$&
  $\mathit{rust\text{-}expr}_{\mathit{bool}}
   \mid P \wedge Q$ \\
  && $\mid\; \texttt{forall}(|x{:}\tau\;\texttt{in}\;(l,h)|\;P)$ \\
  && $\mid\; \texttt{exists}(|x{:}\tau\;\texttt{in}\;(l,h)|\;P)$ \\[3pt]
$Q$ &$::=$& $\cdots \mid \texttt{old}(e)$
  \quad\textrm{\scriptsize(ensures only)} \\[3pt]
$e$ &$::=$& $\mathit{rust\text{-}expr}_{\mathit{pure}}$
\end{tabular}
\end{center}
Multiple \texttt{requires} and \texttt{ensures} clauses are implicitly
conjoined.  The pseudo-function $\texttt{old}(e)$ evaluates~$e$ in the
pre-state and is valid only within \texttt{ensures} clauses; quantifiers
range over a bounded domain $[l,h)$ (inclusive lower bound,
exclusive upper bound) or, if bounds are omitted, over all
values of type~$\tau$ (with backend implications discussed in
Section~\ref{sec:quantifiers}).

%% Commented out for space
%\paragraph{Semantic note.}
%The grammar above defines the \emph{syntax} of specification
%expressions, but their formal semantics is inherited from Rust's
%expression semantics: $\llbracket P \rrbracket_\sigma$ evaluates~$P$
%as a Rust boolean expression in state~$\sigma$, using Rust's
%standard evaluation rules for arithmetic, comparisons, and method
%calls.  We do not provide an independent mathematical semantics for
%the assertion language; the Hoare triples in this section are
%therefore grounded in Rust's operational semantics as compiled to
%CBMC's GOTO representation.  This is standard practice for
%program-level model checkers~\cite{Chong21,Kroening23} but means
%that the formal story depends on the correctness of the MIR-to-GOTO
%translation (Section~\ref{sec:architecture}).

\paragraph{Validity invariants and safety invariants.}
\label{sec:invariants}
A Rust-specific challenge is that types carry two classes of invariant.
A \emph{validity invariant} $\vInv(\tau)$ must hold at all times (e.g.,
\texttt{bool} $\in \{0,1\}$, valid enum discriminants); the compiler
assumes it unconditionally.  A \emph{safety invariant} $\sInv(\tau)$ is
a semantic property that safe code may assume but the type system does
not enforce (e.g., \texttt{Duration}'s nanosecond field $< 10^9$).
Kani's \texttt{kani::any::<T>()} respects $\vInv(\tau)$ by construction
but does not automatically enforce $\sInv(\tau)$.  The \texttt{Invariant}
trait allows types to express $\sInv(\tau)$ programmatically; when a
\texttt{requires} clause includes an \texttt{Invariant} check, inputs
are constrained accordingly.  We write $\mathit{Arb}(\tau)$ for the set
of values producible by \texttt{kani::any::<T>()}.
%% Commented out for space — enum Arbitrary derivation detail
%For enums, Kani's automatic \texttt{Arbitrary} derivation generates one
%nondeterministic variant index per declared variant and constructs each via
%MIR's \texttt{AggregateKind::Adt}, which assigns the correct discriminant
%regardless of whether the representation is contiguous or custom
%(e.g., \texttt{\#[repr(u8)]} with non-contiguous values like \texttt{A = 10,
%B = 20}).  The selection index and the discriminant encoding are
%decoupled: for example, given \texttt{enum E \{ A = 10, B = 20 \}},
%the selection index ranges over $\{0, 1\}$ (choosing which variant to
%construct), while the discriminant values 10 and 20 are assigned by MIR
%during construction.  The nondeterminism selects which variant to construct; MIR
%semantics ensure the resulting value has a valid discriminant.

\paragraph{Semantic domains.}
We write $\sigma \in \Sigma$ for program states (mappings from locations to
values), $f(\bar{x})$ for a function with parameters~$\bar{x}$ and
body~$B_f$, and $\mathit{Mod}(f) \subseteq \mathit{Loc}$ for the write set
of~$f$.  We write $\sigma[L \leftarrow \mathit{havoc}_\tau]$ for the state
obtained by \emph{havocking}~$L$: replacing every location in~$L$ with a
nondeterministic value drawn from $\mathit{Arb}(\tau)$, thereby
preserving $\vInv(\tau)$ while leaving $\sInv(\tau)$ unconstrained
unless the precondition specifies otherwise.%  (The term originates
%in model checking, where ``havoc'' means ``assign an arbitrary
%value.'')

% ---------------------------------------------------------------------------
\subsection{Function Contracts}
\label{sec:function-contracts}
% ---------------------------------------------------------------------------

A function contract is a triple
$\langle \mathit{Pre},\, \mathit{Post},\, \mathit{Mod} \rangle$ where
$\mathit{Pre}(\bar{x})$ is the precondition,
$\mathit{Post}(\bar{x}, r)$ the postcondition over the return value~$r$,
and $\mathit{Mod}$ the write set.  All contract expressions must be
deterministic and side-effect free (enforced at compile time).
In the postcondition, bare parameter
names refer to their \emph{post-state} values; the \texttt{old($e$)}
construct (described below) is required to reference pre-state values.
For parameters passed by value, the post-state and pre-state values
coincide (the caller's copy is unaffected); for parameters passed by
mutable reference (\texttt{\&mut T}), the post-state value reflects
modifications made through the reference.
The contract establishes the Hoare
triple:
\begin{equation}\label{eq:hoare-contract}
  \bigl\{\ \mathit{Pre}(\bar{x})\ \bigr\}\quad
  r \gets f(\bar{x})
  \quad\bigl\{\ \mathit{Post}(\bar{x}, r)
    \;\wedge\; \mathit{frame}(\sigma, \sigma', \mathit{Mod})\ \bigr\}
\end{equation}
where $\mathit{frame}(\sigma, \sigma', \mathit{Mod})$ asserts that every
location outside $\mathit{Mod}$ is unchanged between the
pre-state~$\sigma$ and post-state~$\sigma'$.  In our running example, the
GCD contract has $\mathit{Pre} \equiv (x > 0 \wedge y > 0)$ and
$\mathit{Post} \equiv (r > 0)$; the write set is empty since
\texttt{gcd} is a pure function.  The loop invariant establishes
the stronger property that the set of common divisors is preserved,
but the postcondition is simplified to avoid intractable quantifier
instantiation over nonlinear bitvector arithmetic; see
Section~\ref{sec:example}.
In practice, contract precision involves a trade-off: a precondition
that is too strong prevents the contract from being applied at call
sites that do not establish the full condition, while a postcondition
that is too weak may leave call-site verification unable to prove
the desired property.  Developers must balance precision against
annotation effort.

\paragraph{Contract checking.}
A contract is verified in a harness annotated with
\texttt{\#[kani::proof\_for\_contract($f$)]}.  Kani instruments the
harness to establish~(\ref{eq:hoare-contract}):
\begin{enumerate}
\item construct arbitrary inputs $\bar{x}$ via $\mathit{Arb}(\tau_i)$,
      respecting Rust's ownership invariants (e.g., no two \texttt{\&mut T}
      arguments alias the same memory);
\item assume $\mathit{Pre}(\bar{x})$;
\item symbolically execute $r \gets f(\bar{x})$;
\item assert $\mathit{Post}(\bar{x}, r)$; and
\item verify $\mathit{writes}(f) \subseteq \mathit{Mod}$ via CBMC's
      frame condition checker.
\end{enumerate}
The instrumented harness is discharged by BMC~(\ref{eq:bmc}): CBMC searches
for an execution that satisfies the precondition but violates the
postcondition or frame condition.  If loop contracts are also present, the
loop body executes only twice in a single BMC query
(c.f. Section~\ref{sec:loop-contracts}), decoupling verification time from
iteration count.

\paragraph{Contract replacement.}
Once~(\ref{eq:hoare-contract}) is established, the contract soundly
replaces $f$ at call sites via
\texttt{\#[kani::stub\_verified($f$)]}, applying a \emph{contract
abstraction rule} that combines the verified contract with frame
reasoning:
\begin{equation}\label{eq:replace}
\frac{\displaystyle
  \bigl\{\ \mathit{Pre}\ \bigr\}\;
  r \gets f(\bar{x})\;
  \bigl\{\ \mathit{Post} \wedge \mathit{frame}\ \bigr\}
}{\displaystyle
  \bigl\{\ P \wedge \mathit{Pre}\ \bigr\}\;
  r \gets f(\bar{x})\;
  \bigl\{\ P|_{\overline{\mathit{Mod}}} \wedge \mathit{Post}\ \bigr\}
}
\end{equation}
where $P$ is the caller's context and $P|_{\overline{\mathit{Mod}}}$
retains the conjuncts of~$P$ that do not depend on locations in
$\mathit{Mod}$; conjuncts that do are discarded as part of the
over-approximation.  The rule is admissible, justified by the
BMC-verified premise: the replacement body havocs $\mathit{Mod}$
and then assumes $\mathit{Post}$, which narrows the nondeterministic
values to those satisfying the postcondition.  No
computation of~$f$ is performed, so $f$'s internal loops need not be
unrolled at call sites.  In the \texttt{TokenBucket} verification of
Section~\ref{sec:example}, this reduction drops verification of the caller
from hours (i.e., the time required to unroll all 64-bit GCD iterations in
context) to seconds, while providing an unbounded guarantee over all
64-bit inputs.

\paragraph{History expressions.}
The \texttt{old}($e$) construct captures the pre-state value of~$e$
for two-state postconditions, essential when a function receives
\texttt{\&mut T}.  For example,
\texttt{ensures(|r| *v == old(*v) + 1)} asserts that the referent
was incremented by exactly one.  Kani rewrites
$\texttt{ensures}(|r|\; Q(\texttt{old}(e), \bar{x}, r))$ into
$\{v \gets \llbracket e \rrbracket_\sigma\}\;
r \gets f(\bar{x})\; \{Q(v, \bar{x}, r)\}$,
where $\sigma$ is the pre-state.

\paragraph{Write sets and havocking.}
For functions taking \texttt{\&mut T}, Kani infers $\mathit{Mod}$
from the signature: all memory reachable through mutable references.
The havoc uses \texttt{kani::any()} (preserving $\vInv(\tau)$), and
the \texttt{modifies} macro converts references to raw pointers
\emph{after} borrow checking via \texttt{untracked\_deref}.
Explicit \texttt{modifies} clauses narrow the inferred set when the
reachable state is too large, improving solver completeness.
For raw pointers, \texttt{modifies} is required: Kani cannot infer
which memory is reachable through a raw pointer, so without an
explicit clause the write set defaults to empty and any write
triggers a DFCC verification failure.
The frame condition in~(\ref{eq:hoare-contract}) assumes that
locations inside and outside $\mathit{Mod}$ do not alias.  Rust's
borrow checker guarantees this for safe code; for \texttt{unsafe}
code, the user must ensure $\mathit{Mod}$ is closed under aliasing.

\paragraph{Recursive functions.}
Contracts support unbounded verification of \emph{self-recursive} functions
via induction on call depth, implementing the standard proof
rule~\cite{Apt09}:
\begin{equation}\label{eq:recursion}
\frac{\displaystyle
  \bigl\{\ \mathit{Pre}\ \bigr\}\; f(\bar{x}) \;\bigl\{\ \mathit{Post}\ \bigr\}
  \;\vdash\;
  \bigl\{\ \mathit{Pre}\ \bigr\}\; B_f \;\bigl\{\ \mathit{Post}\ \bigr\}
}{\displaystyle
  \vdash \bigl\{\ \mathit{Pre}\ \bigr\}\; f(\bar{x}) \;\bigl\{\ \mathit{Post}\ \bigr\}
}
\end{equation}
The premise assumes the contract holds for recursive calls and
verifies the body under that assumption.
Rule~(\ref{eq:recursion}) is sound for partial correctness unconditionally,
and for total correctness provided the recursion terminates under a
well-founded measure.  \emph{Kani does not currently verify termination for
recursive functions}; termination remains the user's responsibility.
Mutual recursion ($f \to g \to f$) is not supported; Kani emits a
compile-time error when it detects a contracted function involved
in a mutually recursive cycle.

\paragraph{Soundness and panic semantics.}
Kani interprets Hoare triples under \emph{partial correctness with
panic safety}: $\{P\}\;S\;\{Q\}$ means that if $P$ holds and $S$
terminates normally (without panicking), then $Q$ holds.  Panics
(from \texttt{assert!}, \texttt{unwrap()}, arithmetic overflow, or
out-of-bounds access) are translated into assertion violations and
are therefore checked as part of the verification.  A successful
Kani run guarantees both the absence of panics and the
postcondition, for all inputs satisfying the precondition, up to
the following assumptions: harness coverage of reachable inputs,
aliasing closure of $\mathit{Mod}$ for \texttt{unsafe} code,
user-guaranteed termination, sequential execution
(no threads or \texttt{async}), no Stacked
Borrows~\cite{Jung20} / Tree Borrows~\cite{Villani25}
pointer-aliasing modeling (the primary class of UB in
\texttt{unsafe} Rust), and FFI calls executing outside CBMC's
memory model unless stubbed (\S\ref{sec:stubbing}).
The last two are the most significant in practice.
Miri~\cite{Jung26miri} checks aliasing violations dynamically,
making it a natural complement.

% ---------------------------------------------------------------------------
\subsection{Loop Contracts}
\label{sec:loop-contracts}
% ---------------------------------------------------------------------------

For programs where a completeness threshold exists but is too costly to
reach, such as the 64-bit GCD loop, loop contracts provide an alternative
path to unbounded verification.  They replace exhaustive loop unrolling with
an application of the Floyd--Hoare \textsc{While} rule.

A loop contract consists of three clauses:
\begin{itemize}
\item \texttt{loop\_invariant($I$)},
\item \texttt{loop\_modifies($W$)}, and
\item \texttt{loop\_decreases($d$)}.
\end{itemize}
\noindent
For a
loop \texttt{while}~$G$~$\{B\}$ with post-loop continuation~$C$, the proof
obligation is:
\begin{equation}\label{eq:while}
\frac{\displaystyle
  \bigl\{\ I \wedge G\ \bigr\}\; B \;\bigl\{\ I\ \bigr\}
  \qquad \mathit{writes}(B) \subseteq W
}{\displaystyle
  \bigl\{\ I\ \bigr\}\;
  \texttt{while}\ G\; \{B\}\;
  \bigl\{\ I \wedge \neg G \wedge \mathit{frame}(\sigma,\sigma',W)\ \bigr\}
}
\end{equation}
Kani verifies rule~(\ref{eq:while}) through a two-level transformation
that produces a single BMC query covering all three proof obligations.

%\paragraph{Two-level architecture.}
At the MIR level, Kani restructures the loop CFG (desugaring
\texttt{for} loops and hoisting variables for nesting).  At the GOTO
level, CBMC's DFCC pass rewrites the loop into four blocks executed
in a single BMC run: Prehead (snapshots and initialization), Head
(guard evaluation and body execution), Step (base-case assertion,
havoc~$W$, assume~$I$), and Exit (frame assumption).  The control
flow is Prehead $\to$ Head (base case) $\to$ Step $\to$ Head
(inductive step) $\to$ cut.  The loop body executes exactly twice,
making verification time independent of the iteration count.

%\paragraph{Design choices.}
Two aspects of this transformation are non-obvious.  First, the Step
block sits \emph{between} the two passes through Head: if the loop
guard is false on entry, Step is never reached and the havoc never
fires, correctly preserving the program state for zero-iteration
loops.  Second, the inductive step routes through the actual compiled
guard in Head rather than synthetically assuming it, so the body
executes under $I \wedge G$ with all safety checks on the guard
expression intact.

%% Commented out for space
%\paragraph{Vacuous-loop handling.}
%The placement of the Step block \emph{between} the two passes through
%Head, rather than before the first, is a deliberate design choice.  If
%the loop guard is false on entry, the flow goes Prehead $\to$ Head $\to$
%Exit: the Step block is never reached, the havoc never fires, and the
%program state passes through unmodified.  A na\"{\i}ve transformation
%that havocked before testing the guard would destroy the concrete state
%of a zero-iteration loop, replacing it with nondeterministic values
%constrained only by the invariant and introducing spurious
%counterexamples for code following the loop.
%
%\paragraph{Guard routing.}
%In the inductive step, the flow goes Step (havoc, assume~$I$) $\to$
%Head (evaluate~$G$).  The guard is evaluated in a state where the
%invariant holds but has not been conjoined with the guard itself.  If
%$G$ is true, the body executes under $I \wedge G$, which provides the
%strict condition needed for operations guarded by the loop condition
%(e.g., an array access at index~$i$ where the invariant gives
%$i \leq n$ but the guard provides $i < n$).  Routing through the
%actual compiled guard avoids duplicating the guard expression, which
%at the GOTO level may involve temporary variables, pointer arithmetic,
%and safety-relevant operations that would need their own checks.

%\paragraph{Nested loops and composition.}
Once an inner loop's contract is verified, its abstraction (i.e., havoc
$W_{\mathit{in}}$, assume $I_{\mathit{in}} \wedge \neg
G_{\mathit{in}}$) over-approximates its concrete behavior.  The
outer loop's inductive step therefore remains sound against the
abstracted inner body, provided $W_{\mathit{in}} \subseteq
W_{\mathit{out}}$.  Function-level composition follows the same
argument: all loop abstractions are applied before checking the
function's postcondition.

In our running example, the GCD loop invariant asserts that $a > 0$ and
that the set of common divisors of $(x, y)$ equals the set of common
divisors of $(a, b)$ at every iteration.  The base case holds trivially
($(a, b) = (x, y)$ on entry); the inductive step follows because Euclid's
step preserves the set of common divisors; and the abstraction delivers
$\neg G \equiv (b = 0)$, from which the postcondition $a > 0$
follows directly from the invariant's first conjunct.  The
invariant also implies $a \mid x \wedge a \mid y$ at this point,
but verifying that implication automatically requires quantifier
instantiation that current SMT solvers cannot perform; see
Section~\ref{sec:example}.

%\paragraph{For-loop support.}
Rust's \texttt{for} loops desugar into iterator protocol calls with opaque
internal state that CBMC cannot inspect.  Kani rewrites annotated
\texttt{for} loops into \texttt{while} loops over a simplified
\texttt{KaniIntoIter} abstraction that exposes only the current index and
the iteration bounds.  This allows invariants to be stated over the loop
index rather than private iterator fields.  The rewrite is exact for
range-based \texttt{for} loops; for custom iterators, the user must provide
loop-level contracts that reason about the iterator's abstraction.

\paragraph{Termination.}
Without \texttt{loop\_decreases}, rule~(\ref{eq:while}) establishes only
partial correctness: the invariant is checked at entry and preserved by each
iteration, but non-termination is not ruled out.  Kani emits a warning when
\texttt{loop\_decreases} is absent.  A \texttt{loop\_decreases($d$)} clause
adds the termination proof obligation:
\[
  \bigl\{\ I \wedge G \wedge d = d_0 \wedge d_0 \geq 0\ \bigr\}\;\; B
  \;\;\bigl\{\ 0 \leq d < d_0\ \bigr\},
\]
upgrading the result to total correctness.  The measure~$d$ is restricted
to non-negative integer expressions; more general well-founded orders
(e.g., lexicographic tuples) are not yet supported.  In our example,
\texttt{loop\_decreases(b)} is discharged because each iteration replaces
$b$ with $a \bmod b < b$, and $b \geq 0$ is preserved.  Currently, the
underlying CBMC engine supports only simple integer expressions in
\texttt{decreases} clauses; struct field projections and tuple-based
lexicographic measures are not yet supported.% and are silently ignored by
%CBMC's instrumentation pass.\footnote{\url{https://github.com/model-checking/kani/issues/3168}}

% ---------------------------------------------------------------------------
\subsection{Quantifiers}
\label{sec:quantifiers}
% ---------------------------------------------------------------------------

Kani provides first-order quantifiers as procedural macros:
\begin{lstlisting}
kani::forall!(|x: T in (lo, hi)| P(x))
kani::exists!(|x: T in (lo, hi)| P(x))
\end{lstlisting}
These compile to CBMC-level quantified expressions.  With SAT solvers
(the default), CBMC eagerly expands quantifiers over the specified
range, so bounds must be compile-time constants; Kani emits a warning
when the range exceeds 1000 values.  With SMT backends
(\texttt{--solver z3}, \texttt{cvc5}, or \texttt{bitwuzla}),
quantifiers are passed directly to the solver as first-order formulas,
supporting runtime-valued bounds.
%% Commented out for space — slice postcondition example
%Quantifiers are particularly valuable in postconditions over slices.  For
%example, asserting that every byte of an output buffer is nonzero requires:
%%
%\begin{lstlisting}
%#[kani::ensures(|result: &[u8]|
%    kani::forall!(|i: usize in (0, result.len())|
%        result[i] != 0))]
%\end{lstlisting}
%%
%Without quantifiers, such a property would require a specification loop,
%which would itself need an invariant or an unwinding bound, reintroducing
%exactly the scalability problem that contracts are meant to solve.  When
%used with the SAT backend, quantifier bounds must be literal constants;
%using \texttt{result.len()} as an upper bound requires the SMT backend.

% ---------------------------------------------------------------------------
\subsection{Function Stubbing}
\label{sec:stubbing}
% ---------------------------------------------------------------------------

Function stubbing replaces a function's implementation with a handwritten
mock for the duration of a specific harness, specified via the
\texttt{\#[kani::stub($f$, $g$)]} attribute.  The replacement is a
MIR-to-MIR transformation: when the compiler requests the MIR for~$f$,
Kani returns the MIR for~$g$ instead.  Kani validates signature
compatibility (matching parameter types, return type, and generic
arity) at compile time; without this check, a stub could silently operate
on differently typed values, producing spurious proofs.

Stubbing serves three purposes: (1)~replacing functions with features
unsupported by the verification backend (inline assembly, FFI); (2)~replacing
expensive implementations with lightweight nondeterministic abstractions;
and (3)~environmental modeling, where system interactions such as clock
reads or random number generation are replaced with unconstrained
nondeterministic values.

\paragraph{Stubs versus verified contracts.}
Plain stubs are \emph{not} checked against the original implementation: the
user asserts that $g$ is a sound model of~$f$, and Kani trusts this claim.
A verification result that relies on unverified stubs is therefore
\emph{conditional}: it is sound only if every stub faithfully
over-approximates the behavior of the function it replaces.
Verified contracts (Section~\ref{sec:function-contracts}), by contrast,
machine-check the contract against the implementation before admitting the
abstraction; subsequent uses of \texttt{stub\_verified} are then provably
sound by rule~(\ref{eq:replace}).  The two mechanisms are complementary:
contracts for formally established behavioral abstractions, stubs for
environmental modeling where no implementation exists to verify against.
In the \texttt{TokenBucket} example of Section~\ref{sec:example}, the
\texttt{gcd} function uses a verified contract, while
\texttt{Instant::now}, a system call whose semantics Kani cannot
model, uses a plain stub that returns a nondeterministic \texttt{Instant}.
Together, they reduce the verification of \texttt{TokenBucket::new} from
an intractable loop-unrolling problem to a bounded check with
an unbounded correctness guarantee.

\section{Evaluation}
\label{sec:eval}

We evaluate Kani through three research questions grounded in the concerns of
industrial practitioners:

\begin{description}
  \item[\textbf{RQ-I}] Does Kani find real bugs that testing and fuzzing miss?
  \item[\textbf{RQ-II}] Do function and loop contracts extend bounded verification
    to stronger, unbounded guarantees in practice?
  \item[\textbf{RQ-III}] Is Kani sustainable in continuous integration across
    diverse projects over time?
\end{description}

\begin{table}[htbp]
\centering
\footnotesize
\caption{Subset of open-source projects using Kani.}
\label{tab:projects}
\begin{tabular}{@{}llrrl@{}}
\toprule
Project & Domain & Harnesses & Runtime & Contracts \\
\midrule
Firecracker~\cite{Agache20} & Cloud Infrastructure & 34 & 21 min & \checkmark \\
s2n-quic~\cite{s2nquic} & Network Protocol & 102 & 23 min & -- \\
Hifitime~\cite{Hifitime} & Aerospace & 153 & 42 min & \checkmark \\
verify-rust-std~\cite{Cook25verifyrust} & Standard Library & 16,748 & 69 min & \checkmark \\
\midrule
zerocopy~\cite{zerocopy} & Serialization & 10 & 3 min & \checkmark \\
lading~\cite{lading} & Load Testing & 22 & 2 min & -- \\
x86\_64~\cite{x86_64} & Hardware & 6 & {$<$1 min} & -- \\
rust-sel4~\cite{rustsel4} & Microkernel & 1 & 4 min & -- \\
\bottomrule
\end{tabular}
\end{table}

\noindent
Table~\ref{tab:projects} lists representative open-source projects
across multiple domains that use Kani in CI.
Our evaluation focuses on a subset of these:
Firecracker, the virtual-machine monitor behind AWS Lambda~\footnote{\url{https://aws.amazon.com/lambda/}} and AWS Fargate~\footnote{\url{https://aws.amazon.com/fargate/}};
s2n-quic, Amazon's implementation of the IETF QUIC transport protocol~\cite{quic2017};
Cedar~\cite{Cedar24}, an open-source authorization policy language and engine
(which does not use Kani in its regular CI);
Hifitime, a high-precision time management library used in the Moon landing
mission Blue Ghost Mission~1\footnote{\url{https://fireflyspace.com/missions/blue-ghost-mission-1/}};
and the Rust standard library verification campaign~\cite{Cook25verifyrust}.

% ---------------------------------------------------------------------------
\subsection{Bug-Finding Effectiveness}
\label{sec:rq1}
% ---------------------------------------------------------------------------

We highlight three cases that illuminate qualitatively different failure modes.

\begin{table}[htbp]
\centering
\footnotesize
\caption{Bugs found by Kani that testing and fuzzing missed.}
\label{tab:bugs}
\begin{tabular}{lll}
\toprule
Project & Bug type & Missed by \\
\midrule
s2n-quic & \texttt{try\_fit} assertion & Tests, fuzz (16M iter.) \\
s2n-quic & Pkt.\ number decode overflow & Tests only \\
Firecracker & Rate limiter rounding & Tests, fuzz \\
Firecracker & VirtIO panic from guest & Tests, fuzz \\
Cedar & \texttt{contains\_at\_least\_two} & Tests, diff.\ testing \\
Hifitime & \texttt{normalize()} sign overflow & Tests \\
Hifitime & \texttt{i64::MIN.abs()} panic & Tests \\
Hifitime & Float multiply NaN loop & Tests \\
Hifitime & Epoch Eq/Ord inconsistency & Tests \\
Hifitime & Duration Eq/Ord zero-crossing & Tests \\
Hifitime & \texttt{is\_gregorian\_valid} overflow & Tests \\
\bottomrule
\end{tabular}
\end{table}

\paragraph{s2n-quic: finding bugs at encoding boundaries.}
Amazon's QUIC implementation, s2n-quic,
uses Kani alongside the Bolero\footnote{\url{https://github.com/camshaft/bolero}} property-testing framework,
allowing each harness to run as both a fuzz test
and a Kani proof with a single attribute.  The most striking result involves
the \texttt{try\_fit} function, which determines how many bytes of data to
fit into a QUIC \texttt{Stream} frame given a packet's remaining capacity, a
subtle calculation because the variable-length integer encoding of the frame
length increases at capacity boundaries.  A differential Bolero harness
compared the implementation against a reference model.  Running the \textit{libfuzzer}
engine for over ten minutes (16,777,216 executions) found no failure.  Kani
found a failing assertion in 20 seconds.  The failure lived near the boundary
between one-byte and two-byte integer encodings, a sparse region of the input
space that coverage-guided fuzzing has no gradient toward; Kani's symbolic
execution reached it directly.

A second s2n-quic case illustrates complementarity: a bug in
\texttt{decode\_packet\_number} was found by Kani in 2.8 seconds and
independently by fuzzing in under a minute.  When both methods find the same
bug, the Kani result is more actionable: the failing check identifies a
specific named property rather than an unexplained crash, and Kani
generates a concrete counterexample (via its \emph{concrete
playback}\footnote{\url{https://model-checking.github.io/kani/reference/experimental/concrete-playback.html}} feature)
that the developer can replay as a standard Rust test.

\paragraph{Firecracker: verifying properties invisible to testing.}
The Firecracker virtual machine monitor applies Kani to two
security-critical components.  For the I/O rate limiter, which uses a
token-bucket algorithm where tokens accumulate over time and each I/O
operation consumes tokens, the property of interest is that a microVM
cannot exceed its configured I/O bandwidth in any one-second interval.
This is inherently time-dependent: the number of tokens replenished
depends on when \texttt{auto\_replenish} is called relative to the system
clock, making it impossible to test exhaustively.
Kani resolves this by stubbing \texttt{libc::clock\_gettime} with a
nondeterministic stub that returns monotonically non-decreasing
\texttt{Instant} values, turning the temporal dimension into a symbolic
variable.  The resulting harnesses found several bugs in the rate-limiter
implementation, the most significant being a rounding error that allowed a
guest to exceed its I/O budget by up to 0.01\% in adversarially timed
invocations.  The error depends on the exact sub-millisecond timing of
replenishment calls relative to the system clock, a condition that testing
cannot control deterministically and that fuzzing cannot target without an
explicit time model.

Firecracker also implements the VirtIO device emulation
protocol~\cite{VirtIO11}, where shared memory between host and guest
creates an adversarial attack surface: an untrusted guest can write
arbitrary values into the shared descriptor and ring buffers.
Kani verified conformance to Section~2.6.7.2 of the VirtIO specification
using nondeterministic guest memory to represent this adversarial scenario.
Kani found an additional bug: a guest could cause
Firecracker to panic on boot by placing a VirtIO queue component's starting
address in the MMIO gap.

\paragraph{Cedar: a logic bug in policy evaluation.}
Cedar is an open-source authorization policy language and engine.  A Kani
experiment on Cedar's evaluator found a bug in the
\texttt{contains\_at\_least\_two} function, which checks whether a string contains at
least two occurrences of a substring.  When called with a multibyte
character as the search query, the function could slice a \texttt{\&str}
across a non-character boundary, resulting in a runtime panic.
The bug was silent under all
existing tests.  Kani's symbolic execution over all possible collection inputs produced
a counterexample that violated the function's intended invariant; the fix was
merged\footnote{More information available at \url{https://github.com/cedar-policy/cedar/pull/1037}.}.
This case is notable because Cedar already
uses differential testing against a formal reference model, yet the bug
survived into production code.

%All primary case study projects run Kani harnesses in CI on every
%code change (Table~\ref{tab:projects}); CI sustainability is
%examined in detail in RQ-III (Section~\ref{sec:rq3}).

\finding{Answer to RQ-I:}{Kani found eleven bugs across four production
Rust codebases that testing and fuzzing missed, including bugs that
survived millions of fuzz iterations (s2n-quic), bugs in
time-dependent code invisible to testing without a time oracle
(Firecracker), and differential testing against a
formal reference model (Cedar).  Beyond bug-finding, the harnesses
that detected these bugs also serve as machine-checked specifications
that persist in CI, foreshadowing the stronger correctness guarantees
explored in RQ-II.}

% ---------------------------------------------------------------------------
\subsection{Contract-Based Unbounded Verification in Practice}
\label{sec:hifitime}
% ---------------------------------------------------------------------------

We present a detailed case study of Kani's specification language
applied to Hifitime, a high-precision time-management
library used in aerospace and astrodynamics applications.  Hifitime stores
durations as a pair $(\mathit{centuries}{:}\,\texttt{i16},\,
\mathit{nanos}{:}\,\texttt{u64})$ where the nanosecond field must satisfy
the normalization invariant $\mathit{nanos} < \mathit{NPC}$ (i.e., the number of
nanoseconds in one century) except at the saturation extremes
\texttt{Duration::MIN} and \texttt{Duration::MAX}.  All arithmetic
operations call \texttt{normalize()} to re-establish this invariant, making
it the library's central safety-critical function.

\subsubsection{Phase I: Panic-Freedom Harnesses}

Hifitime's original Kani integration, developed by the library's author in
2022--2023 with no (substantial) prior formal verification experience, consisted of 11
harnesses checking only \emph{panic-freedom}: that operations do not
trigger arithmetic overflows, null dereferences, or out-of-bounds
accesses.  Despite the absence of
postconditions, this baseline found 8 bug categories in a single pull
request\footnote{More information available at \url{https://github.com/nyx-space/hifitime/pull/192}.}, including overflow at boundary values of
\texttt{normalize()}, sign errors in \texttt{neg()}, and
overflow in \texttt{PartialEq} and arithmetic operations near
\texttt{Duration::MIN} and \texttt{Duration::MAX}, all of which survived
78 existing tests.

By 2024 the harness count had grown to 168 through a
combination of manual and automated harness generation.
This expansion revealed a fundamental scalability limit: of 158 automatically
generated harnesses, 57 timed out at the 60-second CI budget.  The root cause
was that bounded model checking re-verified the \texttt{normalize()} loop
inside every caller, with the loop's nanosecond adjustment requiring many
unrolling iterations in the worst case.

\subsubsection{Phase II: Contract-Based Functional Verification}

To address both the scalability problem and the expressive gap, we upgraded
the Hifitime verification using Kani's function and loop contracts.  This
phase was conducted using an AI coding assistant,
providing a concrete data point on AI-assisted formal specification.

The central idea is compositional verification via contracts.  We
expressed the normalization invariant ($\mathit{nanos} < \mathit{NPC}$)
as a postcondition on \texttt{normalize()}, enabling callers to use
\texttt{stub\_verified} (Equation~\ref{eq:replace}) to replace the
implementation with its contract abstraction i.e. eliminating the need to
re-unroll the internal loop at each call site.  Building on this, we
annotated 84 functions with \texttt{ensures} postconditions and 41 with
\texttt{requires} preconditions, converted 12 of the existing 168
panic-freedom harnesses into \texttt{proof\_for\_contract} harnesses
that verify functional properties, and added 61 new
\texttt{proof\_for\_contract} harnesses and 80 standalone proofs with
explicit functional assertions.  This compositional approach enabled
us to go beyond panic-freedom and prove six new classes of functional
properties, from normalization invariants and algebraic laws to
encode--decode identities and specification consistency
(see Table~\ref{tab:hifitime-contracts}).
\begin{table}[htbp]
\centering
\footnotesize
\caption{Classes of functional properties proved with contracts in Hifitime (Phase~II).}
\label{tab:hifitime-contracts}
\begin{tabular}{@{}llr@{}}
\toprule
Property class & Representative harnesses & Max time \\
\midrule
Normalization &
  \texttt{normalize}: $\mathit{nanos} < \mathit{NPC}$;
  idempotence &
  0.2s \\
Algebraic laws &
  Commutativity of $+$; identity;
  $a - a = 0$ &
  1.1s \\
Encode--decode &
  \texttt{to\_parts}/\texttt{from\_parts};
  epoch add/subtract &
  4.5s \\
Calendar validity &
  Gregorian ranges; leap-year rule &
  16.4s \\
Loop termination &
  \texttt{Mul<f64>}; \texttt{is\_gregorian\_valid} (4 invariants) &
  13.6s \\
Spec.\ consistency &
  Eq/Ord agreement &
  0.5s \\
\midrule
\textbf{Total} & \textbf{153 proofs (73 contract + 80 standalone)} & \\
\bottomrule
\end{tabular}
\end{table}

For loops that require unbounded reasoning, Kani's loop contracts
(\S\ref{sec:loop-contracts}) establish guarantees independent of
iteration count.  As an example, the \texttt{Duration::Mul<f64>}
precision loop iteratively scales a multiplier \texttt{new\_val}
(computed as \texttt{q * 10\^{}p}) to find the required decimal
precision.  We annotated it with a loop invariant and a decreases
clause:
\begin{lstlisting}
let mut p: i32 = 0;
let mut new_val: f64 = q;
...
#[kani::loop_invariant(p >= 0 && p <= 19)]
#[kani::loop_decreases(19i32.wrapping_sub(p))]
while new_val.is_finite()
    && (new_val.floor() - new_val).abs() >= f64::EPSILON
    && p < 19
{
    p += 1;
    new_val = q * ten.powi(p);
}
\end{lstlisting}
The invariant is inductive and the decreases clause proves
termination, together establishing \emph{total correctness}: the loop
terminates with \texttt{p~$\in$~[0,\,19]} for all \texttt{f64}
inputs, with no assumptions on~\texttt{q}.  In isolation, Kani
verifies the loop contract in 0.2\,s; without it, CBMC unrolls 19
iterations and takes 91.8\,s.  In the full crate context,
verification of \texttt{Mul<f64>} completes in under 70\,s with the
contract, compared to intractable without it.

The result is 153 active harnesses (up from 168 in Phase~I), of
which all prove functional correctness properties beyond
panic-freedom.  The most consequential change from Phase~I is in the
\emph{class of guarantee}: Phase~I proves that operations do not
crash; Phase~II proves that they compute correctly.
The timeout problem that
surfaced in 2024 (57 harnesses exceeding the CI budget) is addressed
by contracts, which allow callers to reason about
\texttt{normalize()} via its specification rather than re-verifying
its internal loop.

\textit{Bugs found by contracts.}
As a side effect of establishing these functional properties, Kani
uncovered six previously unknown bugs, properties that the original
panic-freedom harnesses could not express.

\textbf{\texttt{(i) total\_nanoseconds()} sign error.}  The implementation
  used subtraction instead of addition for durations spanning more than one
  century in the negative direction, producing incorrect values for any
  \texttt{Duration} with \texttt{centuries} $< -1$.
  The encode--decode identity
  (\texttt{from\_total\_nanoseconds(d.total\_nanoseconds()) == d})
  condition immediately falsified the postcondition, providing a concrete
  counterexample.  The fix is a one-character source change.  Notably,
  one existing integration test was \emph{asserting the buggy behavior}
  and had to be corrected alongside the implementation.  The
  postcondition was derived from the \texttt{Duration} type's encoding
  invariant (documented in the struct definition), providing an
  independent specification against which both the implementation and
  the test were checked.  The test oracle, having been written against
  the implementation's output, encoded the same error.

\textbf{(ii) \texttt{i64::MIN.abs()} arithmetic panic.}
  \texttt{Unit::Mul<i64>} called \texttt{.abs()} on \texttt{i64::MIN},
  which panics because the result does not fit in \texttt{i64}.  The
  normalization contract found this; the fix uses
  \texttt{.unsigned\_abs()}.

\textbf{(iii) NaN in float multiply.}  When \texttt{new\_val} overflows
  to infinity in the \texttt{Mul<f64>} precision loop, the loop
  invariant (\S\ref{sec:hifitime}) falsified on the overflow
  iteration.  Two source-level guards were added.

\textbf{\texttt{(iv) Epoch::PartialEq} and \texttt{Ord} inconsistency.}
  The method \texttt{Duration::PartialEq} deliberately treats opposite-sign values
  as equal (representing the same interval length regardless of
  direction), but \texttt{Epoch::PartialEq} delegated to this
  implementation.  Meanwhile, \texttt{Epoch::Ord} used a lexicographic
  comparison that is sign-aware.  The Rust standard library requires
  that \texttt{PartialEq} and \texttt{Ord} agree: \texttt{a == b} must
  imply \texttt{a.cmp(b) == Ordering::Equal}.  This invariant was
  violated for any pair of epochs differing in sign.

\textbf{\texttt{(v) Duration::PartialEq} and \texttt{Ord} inconsistency
  (zero-crossing).}
  \texttt{Duration::PartialEq} has a custom zero-crossing special case
  that treats opposite-sign durations as equal, but \texttt{Duration::Ord}
  is derived (lexicographic on fields) and does not.  Two durations can
  therefore satisfy \texttt{a == b} and \texttt{a < b} simultaneously,
  violating Rust's \texttt{Eq}/\texttt{Ord} contract.  This bug was
  discovered during the investigation of Bug~4: tracing the
  \texttt{Epoch} inconsistency to its root cause in \texttt{Duration}.
  %The maintainer confirmed it as a
  %bug\footnote{More information available at \url{https://github.com/nyx-space/hifitime/issues/469}.}.

\textbf{(vi) \texttt{is\_gregorian\_valid} overflow.}
  The Gregorian calendar validity check overflows at
  \texttt{year == i32::MAX} due to an unchecked addition in the
  leap-year computation.  The contract on the calendar validation
  function detected this boundary
  case.\footnote{\url{https://github.com/nyx-space/hifitime/issues/475}}

Bugs~(iv) and~(v) are the most semantically significant findings because they
violate a requirement of the Rust standard library itself, not
merely a domain-specific invariant.  Bug~(iv) was invisible to all
existing tests because no test compared an epoch to its
sign-negation, a conceptually unusual operation for a ``point in
time.''  Kani's symbolic execution generated this input
automatically by exploring all possible \texttt{Epoch} pairs.  The
\texttt{contract\_epoch\_eq\_symmetric} harness expressed the
consistency requirement as a postcondition and immediately produced
a counterexample.  The fix rewrites both \texttt{Epoch::PartialEq}
and \texttt{Epoch::Ord} to use \texttt{total\_nanoseconds()} for a
sign-aware scalar comparison.  This case illustrates why
exhaustive symbolic analysis over all inputs matters: the failing
input lies in a region of the state space that no human tester would
explore, yet it represents a real semantic violation.

\textit{AI-assisted formal specifications.}
The Phase~II specifications were developed using an AI coding
assistant with shell access to the codebase.  Because Kani's
specification language is an extension of Rust itself (i.e., contracts are
Rust boolean expressions; harnesses are Rust functions), the
assistant could leverage its existing Rust knowledge to produce
specifications without learning a separate formal language.  The
workflow was a tight loop: the assistant proposed contract
annotations, wrote them into the source files, ran \texttt{cargo
kani}, observed the result (i.e., success, failure, timeout, or compile
error), and refined the specification in response.  The human
researcher directed which functions to target, provided domain
context, and reviewed the proposed contracts for correctness.

Crucially, the specifications do not need to be trusted: Kani
verifies every contract against the implementation.  This was
demonstrated concretely when three AI-generated contracts failed
verification, and Kani caught all three:
(1)~an intractable postcondition that called the function under
verification in its own \texttt{ensures} clause, doubling the SAT
formula and causing a solver timeout (the assistant simplified it to
a direct arithmetic expression);
(2)~an overly broad symbolic input (\texttt{kani::any::<Epoch>()})
that included the \texttt{TimeScale::ET} variant, pulling
\texttt{sin()} into the SAT formula and causing a timeout (the
assistant restricted the harness to \texttt{TimeScale::TAI}); and
(3)~a \texttt{const fn} incompatibility where Kani's macro expansion
generated non-const code, producing a compile error (the assistant
switched to a standalone proof harness).
These errors classify as one specification design error, one harness
design error, and one Kani limitation.  In all three cases, the
assistant diagnosed the failure from Kani's output and proposed a
fix within the same session.

%The total specification effort comprised 83 \texttt{ensures}
%annotations, 30 \texttt{requires} annotations, and 1
%\texttt{loop\_invariant}, plus 70 \texttt{proof\_for\_contract}
%harnesses and 14 standalone proof harnesses.
%This workflow produced
%two simultaneous outcomes: five implementation bugs discovered
%because the AI-generated postconditions were precise enough to
%falsify incorrect code, and three specification errors caught by
%Kani before they entered the repository.

\finding{Answer to RQ-II:}{Contracts extend bounded panic-freedom
guarantees to stronger functional properties with unbounded coverage.
In Hifitime, 153 functional proofs covered six new classes of functional properties.
Contracts enabled compositional reasoning
and uncovered six
bugs that panic-freedom checks could not express, including two
violations of Rust's own \texttt{Eq}/\texttt{Ord} contract.
% An AI coding assistant accelerated the specification process, with
% Kani serving as the arbiter that caught all three specification
% errors before they entered the repository.
}

% ---------------------------------------------------------------------------
\subsection{Continuous Integration Sustainability}
\label{sec:rq3}
% ---------------------------------------------------------------------------

All projects listed in Table~\ref{tab:projects} run Kani in CI,
and several have done so for years: Firecracker integrated Kani
in 2022~\cite{Chong20}, and s2n-quic and Hifitime followed shortly
after.  This sustained adoption builds on the code-level model
checking methodology established by Chong et
al.~\cite{Chong20,Chong21} for C codebases at AWS, which Kani
extends to Rust.  The practical CI feasibility of Kani has been
demonstrated at scale: the Rust standard library verification
campaign runs 16,748 Autoharness-generated harnesses via
parallelization and incremental caching, achieving a
3.97$\times$ speedup over a na\"{\i}ve sequential
pipeline~\cite{Cook25verifyrust}.  Across the projects in
Table~\ref{tab:projects}, CI runtimes range from under one minute
(x86\_64) to approximately 69 minutes (verify-rust-std), with most
completing in under 25 minutes, which are practical budgets for pull-request
workflows.

A key enabler of CI adoption is Kani's \texttt{cargo kani} interface, which
integrates identically with \texttt{cargo test}.  In s2n-quic, the Bolero
framework allows the same harness to run as both a fuzz test and a Kani proof
via a single \texttt{\#[cfg\_attr(kani, kani::proof)]} attribute, adding
Kani's exhaustive analysis to existing fuzz infrastructure at zero additional
annotation cost.  The breadth of adoption across domains (from VMMs and
network protocols to aerospace, microkernels, and OS
infrastructure) is evidence that Kani's cargo-native workflow
lowers the barrier to verification across a wide range of industrial
Rust contexts.

\finding{Answer to RQ-III:}{Kani is sustainable in CI across diverse
projects and domains.  All projects in Table~\ref{tab:projects} run
Kani on every code change, with CI runtimes ranging from under one
minute to 69 minutes.  The Rust standard library campaign verifies
over 16,000 harnesses per code change.  The \texttt{cargo~kani}
interface and proof-harness-as-code model enable verification to
evolve alongside the codebase.}

%% ---------------------------------------------------------------------------
\textit{Threats to validity.}
Our case studies are not a random sample: the evaluated projects
self-selected into Kani adoption, potentially over-representing
favorable cases; the Cedar experiment partially mitigates this as a
deliberate evaluation on a project not previously using Kani.
Specifications can be wrong: three AI-generated contracts required
correction (all caught by Kani, not human review).  Harnesses
without loop contracts provide bounded guarantees only; we report
unbounded claims only where contracts or passing unwinding assertions
justify them.  The specification language does not address generics
(contracts are verified per monomorphization), dynamic trait
dispatch, or concurrent execution.

\section{Related Work}
\label{sec:related}

Rust verification is an active area spanning dynamic analysis,
deductive verification, and model checking.  On the dynamic side,
Miri~\cite{Jung26miri} detects undefined behavior at runtime,
including Stacked Borrows~\cite{Jung20} and Tree
Borrows~\cite{Villani25} aliasing violations (the primary class of
UB that Kani does not model; \S\ref{sec:function-contracts}).
Rudra~\cite{Bae21rudra} scales to the full crates.io ecosystem via
over-approximation, finding 264 bugs (76 CVEs).
On the foundational side, RustBelt~\cite{Jung18} and
RefinedRust~\cite{Gaher24} provide machine-checked soundness proofs
in Rocq at the cost of substantial expertise.
Kani's proof-harness approach builds on code-level model checking
for C~\cite{Chudnov18s2n,Chong20,Chong21}, extending it to Rust
with type-aware nondeterminism and contract-based unbounded
verification.

%\paragraph{Rust verifiers.}
Verus~\cite{Lattuada23verus} is the closest tool to Kani in ambition.
It embeds specifications and proofs directly in Rust syntax, using
linear ghost types for ownership reasoning and targeting Z3.  The key
difference is the verification paradigm: Verus requires proof code
(lemma functions, triggers, ghost state), while Kani requires only
contracts and harnesses.  Verus's SOSP~2024 case studies report
approximately 5 lines of proof per line of implementation (6.1K
impl., 31K proof); Kani's Hifitime case study adds 125 contract
annotations to 153 harnesses covering a 9.5K-line library.  Kani is
easier to adopt incrementally (zero annotations yield panic-freedom
checks) but is limited to properties expressible within BMC or
inductive contracts; Verus can express richer properties (e.g.,
complex data-structure invariants with ghost state) at higher
annotation cost.

Other Rust verifiers include
Prusti~\cite{Astrauskas22}, a deductive verifier for safe Rust built
on the Viper infrastructure;
Creusot~\cite{Denis22}, which translates MIR into Why3's WhyML with
prophecy variables for mutable borrows;
Flux~\cite{Lehmann23}, which brings liquid types to Rust with low
annotation overhead;
Aeneas~\cite{Ho22} and hax~\cite{Bhargavan25hax}, which extract Rust
into theorem provers (F*, Lean, Rocq) for foundational or
security-critical verification; and
VeriFast~\cite{Jacobs11verifast}, a separation-logic verifier for
unsafe Rust that handles the most pointer-intensive modules in the
Rust standard library campaign~\cite{Cook25verifyrust}.
Kani differs from these tools by operating on MIR, providing
\texttt{cargo}-integrated workflows, and supporting unbounded
verification through function and loop contracts.

\section{Conclusion}
\label{sec:conclusion}

We presented Kani, an open-source model checker for Rust that
provides both bounded and unbounded verification through a
specification language of function contracts, loop contracts,
quantifiers, and function stubbing.  Our evaluation demonstrates
that Kani finds bugs that testing and fuzzing miss (eleven bugs across
four production codebases), that contracts upgrade bounded
panic-freedom checks to unbounded functional correctness proofs (153
functional proofs in Hifitime), and that verification integrates
sustainably into CI across diverse projects (over 16,000 harnesses
in the Rust standard library campaign).  Model checking
requires minimal proof expertise, yet provides exhaustive guarantees that
testing cannot.  Kani makes this capability accessible to any Rust
developer through \texttt{cargo~kani}.

{\textbf{Data Availability.}}
Kani and all evaluated projects are available as open source;
bug fixes are referenced by pull request number in the text.

{\textbf{Acknowledgements.}}
We thank all open-source contributors. % An AI assistant was used
%for prose editing; all technical content is the sole work of the
%authors.

%\section*{Data Availability}
%Kani is available as open-source software under the Apache~2.0 and MIT
%licenses.
%All evaluated projects are publicly available on GitHub.
%%~\footnote{\url{https://github.com/model-checking/kani}}.
%%All evaluated projects are publicly available:
%%Firecracker,%~\footnote{\url{https://github.com/firecracker-microvm/firecracker}},
%%s2n-quic,%~\footnote{\url{https://github.com/aws/s2n-quic}},
%%Hifitime,%~\footnote{\url{https://github.com/nyx-space/hifitime}},
%%zerocopy,%~\footnote{\url{https://github.com/google/zerocopy}},
%%Cedar,%~\footnote{\url{https://github.com/cedar-policy/cedar}},
%%and the Rust standard library verification
%%campaign.%~\footnote{\url{https://github.com/model-checking/verify-rust-std}}.
%%The proof harnesses and contract annotations described in this paper are
%%committed to the respective repositories.  Bug fixes are referenced by
%%pull request number.
%
%\section*{Acknowledgements}
%We thank all the open-source contributors.
%An AI assistant was used during the preparation of this
%manuscript for prose editing and bibliographic verification.
%All technical content, experimental data, and scientific claims
%are the sole work of the authors, who reviewed and approved the
%final text.

\bibliographystyle{ACM-Reference-Format}
\bibliography{refs}

\end{document}